\documentclass[sigconf,natbib=false]{acmart}
\copyrightyear{2026}
\acmYear{2026}
\setcopyright{none}
\acmConference[ReCode '26]{1st Workshop on Code Translation, Transformation, and Modernization }{April 12--18, 2026}{Rio de Janeiro, Brazil}
\acmBooktitle{1st Workshop on Code Translation, Transformation, and Modernization (ReCode '26), April 12--18, 2026, Rio de Janeiro, Brazil}
\acmDOI{}
\acmISBN{}

\usepackage{glossaries}
\newacronym{MDM}{MDM}{model-driven migration}
\newacronym{TOS}{TOS}{terminal operating system}
\newacronym{ORM}{ORM}{object-relational mapping}
\newacronym{VM}{VM}{virtual machine}
\newacronym{UI}{UI}{user interface}
\newacronym{IDE}{IDE}{integrated development environment}
\newacronym{ESG}{ESG}{environmental, social, and governance}
\newacronym{RQ}{RQ}{research question}
\newacronym{LLM}{LLM}{large language model}

\AtBeginDocument{%
  }

\RequirePackage[
  datamodel=acmdatamodel,
  style=acmnumeric,
  ]{biblatex}

\begin{document}

\title[Model-Driven Legacy System Modernization at Scale]{Model-Driven Legacy System Modernization at Scale}
\thanks{This is the author’s version of the work accepted for publication in the Proceedings of ReCode '26 (ACM). The definitive Version of Record will be published in the ACM Digital Library. This version is provided for personal use and scholarly purposes.}

\author{Tobias Böhm}
\orcid{0009-0003-5409-9384}
\affiliation{%
  \institution{Trier University of Applied Sciences}
  \country{Germany}
}
\email{t.boehm@hochschule-trier.de}

\author{Jens Guan Su Tien}
\orcid{0009-0004-6449-0552}
\affiliation{%
  \institution{Trier University of Applied Sciences}
  \country{Germany}
}
\email{je.tien@hochschule-trier.de}

\author{Mohini Nonnenmann}
\orcid{0009-0001-2440-1285}
\affiliation{%
  \institution{Trier University of Applied Sciences}
  \country{Germany}
}
\email{m.nonnenmann@hochschule-trier.de}

\author{Tom Schoonbaert}
\orcid{0009-0003-0142-1878}
\affiliation{%
  \institution{Euroports}
  \country{Belgium}
}
\email{tom.schoonbaert@euroports.com}

\author{Bart Carpels}
\orcid{0009-0001-8991-7688}
\affiliation{%
  \institution{Euroports}
  \country{Belgium}
}
\email{bart.carpels@euroports.com}

\author{Andreas Biesdorf}
\orcid{0000-0003-0206-7746}
\affiliation{%
  \institution{Trier University of Applied Sciences}
  \country{and Siemens AG, Germany}
}
\email{a.biesdorf@hochschule-trier.de}

\renewcommand{\shortauthors}{Böhm et al.}

\begin{abstract}
This experience report presents a model-driven approach to legacy system modernization that inserts an enriched, technology-agnostic intermediate model between the legacy codebase and the modern target platform, and reports on its application and evaluation. The four-stage process of analysis, enrichment, synthesis, and transition systematically extracts, abstracts, and transforms system artifacts. We apply our approach to a large industrial application built on legacy versions of the .NET Framework and ASP.NET MVC and show that core user interface components and page structures can be migrated semi-automatically to a modern web stack while preserving functional behavior and essential non-functional qualities. By consolidating architectural knowledge into explicit model representations, the resulting codebase exhibits higher maintainability and extensibility, thereby improving developer experience. Although automation is effective for standard patterns, migration of bespoke layout composites remains challenging and requires targeted manual adaptation. Our contributions are: (i)~an end-to-end model-driven process, (ii)~an enriched intermediate model that captures structure, dependencies, and semantic metadata, (iii)~transformation rules that preserve functional behavior and essential non-functional qualities, and (iv)~application and evaluation of the approach in an industrial setting. Overall, model-based abstractions reduce risk and effort while supporting scalable, traceable modernization of legacy applications. Our approach generalizes to comparable modernization contexts and promotes reuse of migration patterns.

\end{abstract}



\maketitle

\section{Introduction}
\label{sec:introduction}
Legacy enterprise systems persist because they are business-critical and operationally mature, yet their accumulated technical debt impedes change and drives maintenance effort~\cite{agilar_systematic_2016,khadka_how_2014}. Manual modernization of very large systems is costly and complex~\cite{deepak_veeravalli_legacy_2022,khadka_how_2014}, motivating model-driven migration (MDM) approaches that introduce an intermediate model to extract, abstract, and transform artifacts in a repeatable way~\cite{barbier_modisco_2010}.

This experience report develops and applies a \emph{model-driven} approach that inserts an enriched, technology-agnostic intermediate model between a legacy codebase and a modern target platform. By separating legacy concerns from target-specific implementation details, the intermediate model increases flexibility in the choice of target technologies. We demonstrate the approach on a large industrial web application built on legacy versions of the .NET Framework and ASP.NET MVC (specific versions are intentionally omitted for confidentiality reasons), showing that core \gls*{UI} components and page structures can be migrated semi-automatically while preserving functional behavior and essential non-functional qualities. Automation is effective for standard patterns; bespoke layout composites remain challenging and require targeted manual adaptation.

Our contributions are: (i)~an end-to-end model-driven process, (ii)~an enriched intermediate model that captures structure, dependencies, and semantic metadata, (iii)~transformation rules that preserve functional behavior and essential non-functional qualities, and (iv)~a documented application and evaluation of the approach in an industrial setting detailing automation boundaries, maintainability effects, and lessons learned.

\paragraph{Research Questions and Paper Structure.}
We investigate how large-scale legacy systems can be migrated semi-automatically to modern technologies by leveraging \gls*{MDM} with an enriched intermediate model. To assess feasibility, we implement and apply the approach to the web-based \glspl*{UI} of a representative legacy application. 

\begin{itemize}
\item[\textbf{RQ1:}] What portions of the migration pipeline can be automated at scale (e.g., extraction of components, mapping of standard UI patterns, code generation), and to which extent?

\item[\textbf{RQ2:}] How can a model-driven migration process centered on an intermediate model be realized for the \gls*{UI} of a large legacy .NET application, and where are the practical boundaries between automation and manual adaptation? 

\item[\textbf{RQ3:}] What effects does the approach have on maintainability and extensibility of the resulting system?

\end{itemize}

Section~\ref{sec:legacy_systems} defines context and scope. Section~\ref{sec:approach} outlines the model-driven approach and pipeline. Section~\ref{sec:case_study} introduces the industrial system. Section~\ref{sec:implementation_results} reports the implementation and results. Section~\ref{sec:evaluation} evaluates findings with respect to the RQs. Section~\ref{sec:limitations_future_work} discusses limitations and future work. Section~\ref{sec:conclusions} concludes.

\section{Legacy Systems in Industry}
\label{sec:legacy_systems}
Although no definitive criterion defines a legacy system, the term generally denotes long-standing, stable production systems that are business-critical and tightly coupled to their organizational environment~\cite{khadka_how_2014}. Such systems resist change and accumulate technical debt. Misalignment with evolving IT strategy that is driven by new capability demands and rising maintenance costs further cements their legacy status~\cite{agilar_systematic_2016,khadka_how_2014}. As a result, many operate in a de facto maintenance mode in which new feature development is heavily constrained by technological limitations~\cite{agilar_systematic_2016}. 

In industrial settings, legacy status is shaped less by technical obsolescence than by business continuity, operational stability, and risk management. While academia often emphasizes outdated technology and maintainability challenges, practitioners prioritize ongoing business value and deep organizational integration~\cite{ agilar_systematic_2016,deepak_veeravalli_legacy_2022, ilieva_enhance_2013,khadka_how_2014}. From this perspective, systems persist as legacy not merely due to age but because they are enduring investments that are costly and risky to replace. This dynamic is reflected in the following recurring characteristics: \\

\textbf{Business criticality} Legacy systems underpin core processes. Failures carry severe consequences. Systems without ongoing business value are often retired before becoming legacy~\cite{khadka_how_2014}.

\textbf{Technological maturity} Built on long-standing, well-under\-sto\-od technology stacks, legacy systems deliver predictable performance, albeit not always optimal by modern standards~\cite{khadka_how_2014}.

\textbf{Limited flexibility} Architectural and technological constraints impede adaptation and integration, slowing time-to-market and innovation~\cite{agilar_systematic_2016,deepak_veeravalli_legacy_2022,ilieva_enhance_2013,khadka_how_2014,ponnusamy_navigating_2023}.

\textbf{High maintenance effort} Outdated technologies, complex architectures, and limited vendor support raise costs, yet replacement risk sustains operation~\cite{agilar_systematic_2016,deepak_veeravalli_legacy_2022,khadka_how_2014}.

\textbf{Knowledge dependence} Expertise is concentrated among few experts. As documentation ages and experts leave, operational risk increases~\cite{khadka_how_2014}.

\textbf{Technology lock-in} Reliance on discontinued platforms and software ecosystems increases migration risk and effort~\cite{deepak_veeravalli_legacy_2022,khadka_how_2014}.

\subsection{Drivers for Modernization}
Modernization is driven by \emph{economic}, \emph{technical}, and \emph{organizational} pressures. High maintenance costs and rising complexity drain resources and limit the ability to adapt~\cite{agilar_systematic_2016,khadka_how_2014}. Technological obsolescence, such as outdated programming languages, architectures, and hardware incompatible with current standards, complicates integration with new systems~\cite{agilar_systematic_2016}. Knowledge erosion due to staff turnover and poor documentation, together with vendor dependence and shrinking tool support, further increases operational risk~\cite{khadka_how_2014}. 

At a \emph{strategic} level, organizations seek agility, interoperability, and scalability as they adapt to cloud, microservices, and API-centric architectures~\cite{adepoju_framework_2024,deepak_veeravalli_legacy_2022,ogunwole_modernizing_2023}. Legacy systems often lack modularity and standardized interfaces, making integration across distributed systems difficult. Modernization therefore aims to re-engineer or migrate systems to architectures that reduce cost and integration effort and improve adaptability~\cite{ilieva_enhance_2013,khadka_how_2014,ponnusamy_navigating_2023}.

\subsection{Legacy Software Modernization} 
Software modernization encompasses the replacement, redevelopment, reuse, or migration of existing software artifacts and platforms~\cite{khadka_how_2014}. In practice, large-scale efforts face several challenges: accumulated technical debt, organizational resistance and user acceptance, and operational and data-related constraints~\cite{ponnusamy_navigating_2023, stavros_challenges_2013}. 

Prior work highlights the need for staged execution, continuous validation, and thorough documentation (e.g.,~\cite{ponnusamy_navigating_2023}). We operationalize these practices with a four-stage pipeline: \emph{analysis} (comprehensive assessment of architecture and artifacts), \emph{enrichment} (explicit semantic documentation in an intermediate model), \emph{synthesis} (rule-based generation with validation at each step), and \emph{transition} (incremental integration and post-deployment maintenance).

Common strategies, such as rehosting, replatforming, re-archi\-tect\-ing, and refactoring, work well for smaller or homogeneous systems. At large scale, they often become impractical due to extensive manual effort and cost~\cite{khadka_how_2014,ponnusamy_navigating_2023}. In particular, heterogeneous \gls*{UI} layers with bespoke composites, internationalization resources, and cross-cutting dependencies resist template-based migration. These limits motivate a technology-agnostic intermediate model that captures structure and semantics to enable repeatable, semi-automatic transformation at scale.

\section{Model-Driven Migration Approach}
\label{sec:approach}
\begin{figure}[htbp]
    \centering
    \includegraphics[width=1\linewidth]{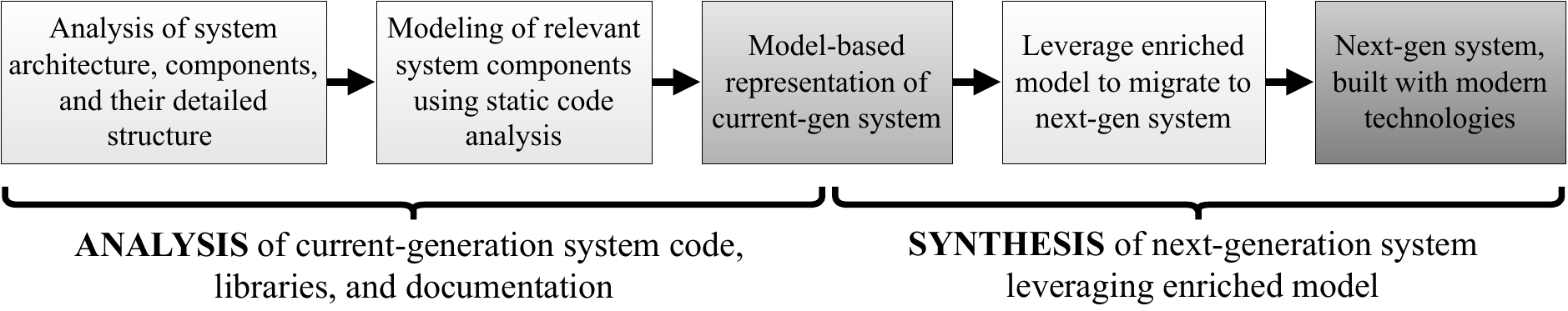}
    \Description[Model-driven migration pipeline]{
    Process diagram: left - analysis of current system (architecture, components, documentation); center - construction of an enriched intermediate model; right - synthesis of a next-generation system using the model.
    }
    \caption{Four-stage model-driven migration pipeline}
    \label{fig:mbm_approach}
\end{figure}

Our approach builds on \emph{model-driven engineering} to enable semi-automatic migration of large-scale legacy systems. It \emph{reduces manual effort} while improving consistency, traceability, and scalability. The approach follows an \emph{iterative pipeline} that maps legacy representations to modern targets via an enriched intermediate model, acting as the pivotal transformation layer (Fig.~\ref{fig:mbm_approach}).

We integrate four steps into a cohesive framework: \emph{Analysis} (Section~\ref{ref:mdm_analysis}) of the current system; \emph{Enrichment} (Section~\ref{ref:mdm_enrichment}) of the intermediate model and abstraction of architectural components; \emph{Synthesis} (Section~\ref{ref:mdm_synthesis}) of next-generation representations; and \emph{Transition} (Section~\ref{ref:mdm_transition}) to the modernized system. Introducing an intermediate model before component migration enables a systematic, incremental pathway and supports reuse of patterns and rules across the system.

The \emph{iterative, technology-agnostic} nature of our approach supports application to other large legacy systems. Migration patterns identified in one subsystem can often be generalized with minor adaptations, reducing time and cost in subsequent efforts. Scaling up introduces governance and tool-interoperability challenges, which we address to maintain consistency and reliability.

\subsection{Analysis of the Current System}
\label{ref:mdm_analysis}

\subsubsection{System Representation}
We begin by constructing an \emph{intermediate model} of the legacy system. The model is populated by structural parsing of source code and configuration artifacts (e.g., projects and build descriptors, service and route definitions, resource files, database schemas), using file-system discovery and, where available, abstract syntax trees. The representation captures (i)~\emph{elements} such as modules, components, interfaces, views, data entities, and resources; (ii)~\emph{typed relations} among them (e.g., containment, depends-on, binds); and (iii)~\emph{stable identifiers and trace links} back to the original artifacts. Normalizing heterogeneous inputs into consistent types and references provides an objective foundation, independent of incomplete or outdated documentation, thereby enabling reliable downstream analysis and transformation. Where dependencies are introduced dynamically (e.g., via reflection or late binding), the extractor records conservative relations and marks them for review during enrichment.

\subsubsection{Architecture and Dependency Extraction}
We extract architecture and dependency information by deriving \emph{typed relations} (e.g., \emph{depends-on}, \emph{calls}, \emph{publishes/subscribes}, \emph{reads/writes}, \emph{binds}, \emph{exposes-endpoint}) between model elements such as modules, components, interfaces, data entities, and services. Relations are inferred from structural sources (e.g., source code, configuration and build descriptors, interface specifications, and database or messaging schemas), and recorded in the intermediate model with provenance and stable identifiers. Initial discovery may require targeted manual reconnaissance, but is \emph{progressively formalized} into reusable extraction rules and queries. Making dependencies explicit enables impact analysis, ordering of transformations, and conformance checks.

\subsection{Abstraction and Model Enrichment}
\label{ref:mdm_enrichment}

\subsubsection{Abstraction of System Components}
To decouple system know\-ledge from platform-specific implementations, we introduce \emph{component abstractions} that normalize heterogeneous artifacts into technology-agnostic model elements. Each abstraction captures (i)~\emph{a stable identity and type} (e.g., Page, View, Component, Service, DataEntity, ResourceString); (ii)~\emph{an explicit interface or contract} (inputs/outputs, bindings, events);~(iii) \emph{essential properties and constraints}; and (iv)~\emph{typed relations to other elements} (e.g., containment, depends-on, binds), with provenance links back to the source artifacts. The abstractions follow three design principles: \emph{minimality} (only information required for transformation and validation), \emph{compositionality} (complex structures are built from simpler elements), and \emph{traceability} (round-trip links to the legacy code and configuration). This normalized view provides a foundation for migration to a modern target platform with reusable transformation rules.

\subsubsection{Semantic Enrichment}
Beyond structure, we annotate model elements with semantics required for safe, automated transformation. Enrichment captures (i)~\emph{intent and role} (e.g., page/view type, endpoint kind); (ii)~\emph{contracts and constraints} (validation rules, required/optional fields, ranges, formats, units); (iii)~\emph{data semantics} (entity and attribute mappings, cardinalities, invariants); (iv)~\emph{interaction and navigation} (routes, deep links, bindings); 
(v)~\emph{internationalization} bindings and locales; and (vi)~\emph{provenance/trace links}. Semantics are inferred from source and configuration artifacts, interface specifications, resource catalogs, tests, and naming/annotation conventions; rules are reusable and produce diagnostic stubs where inference is uncertain. Making implicit knowledge explicit renders it machine-consumable, enabling rule selection, parameterization of generators, conformance checks against invariants, and traceable round-trip links throughout analysis, transformation, and synthesis.

\subsection{Synthesis of Next-Generation System}
\label{ref:mdm_synthesis}

\subsubsection{Transformation Rules}
Using the enriched intermediate model, we define structured, rule-oriented transformation templates for specific page types that map legacy constructs to target-platform equivalents. Each rule specifies a \emph{pattern} over model elements, \emph{guards/preconditions}, a \emph{construction} of target artifacts, and \emph{postconditions/invariants} with trace links for round-trip tracing. Rules cover presentation (pages, navigation), service/API endpoints, data access and persistence, and configuration/infrastructure  and are \emph{parameterized} by target profiles.

Application is \emph{dependency-aware}: rules are executed in phases leveraging contextualization through the model graph; priorities and conflict resolution ensure deterministic outcomes. Rules are designed to be \emph{idempotent} and \emph{incremental}, enabling safe re-runs as the model evolves. Unmatched or low-confidence patterns produce stubs and diagnostics for manual follow-up. This rule-based synthesis yields consistent, repeatable outputs while preserving semantics; edge cases are handled via guards, fallbacks, and assertions that feed into the validation stage.

\subsubsection{Target Platform Analysis}
We characterize the target stack through a machine-readable \emph{target profile} that captures the capabilities, constraints, and idioms of its frameworks, data stores, runtime, and deployment model. The profile enumerates supported constructs and preferred conventions. Transformation rules are informed by this profile via guards and parameters to ensure idiomatic mappings. A fit-gap analysis classifies source patterns as \emph{directly mappable}, \emph{mappable with adaptation}, or \emph{requiring redesign}. The process is iterative: pilots validate mappings, observed deltas refine the profile and rule guards, and re-execution ensures conformance between model semantics and target design principles.

\subsection{Transition to Next-Generation System}
\label{ref:mdm_transition}

\subsubsection{Migration Execution}
We orchestrate migration as an automated, repeatable pipeline driven by the enriched intermediate model, the transformation rules, and the target profile. Generation is \emph{dependency-aware}, \emph{idempotent}, and \emph{incremental}: the pipeline emits target artifacts with stable identifiers and trace links, and runs build, linting, and contract/unit tests as gates. Outputs are classified into \emph{auto-mergeable} changes and \emph{review-required} changes based on rules and invariant checks.

Execution combines several mechanisms according to feasibility and risk: (i) \emph{wrapping/adapters} to integrate selected legacy components behind target interfaces, (ii) \emph{direct code generation} where patterns are fully covered by rules, and (iii) \emph{scaffolding} to support incremental re-implementation for partial matches. Cutover follows proven deployment strategies (e.g., side-by-side “strangler” routing, feature flags, canary or blue-green releases), so that migrated functionality can be validated in production-like conditions with rollback paths. Telemetry and conformance checks verify non-functional objectives which feed back into rule guards and the target profile. This enables safe re-runs of the entire pipeline as the model evolves.

\subsubsection{Integration with Modern Technologies}
Integration proceeds incrementally by wiring generated artifacts to target-platform primitives through configuration and adapters defined in the target profile. Configuration is externalized and managed as code. Compatibility shims isolate residual legacy elements, while modern equivalents are introduced. Observability (logs, metrics, traces) and contract tests verify behavioral and non-functional conformance. This staged integration improves interoperability and performance, reduces residual coupling to legacy infrastructure, and prevents reintroduction of legacy constraints.

\subsubsection{Validation and Feedback}
Validation is embedded throughout the pipeline and targets both technical and business requirements. Model invariants and rule post-conditions act as fail-fast gates; violations surface as diagnostics linked via trace IDs to both legacy artifacts and generated outputs. Feedback is continuous: results update confidence tags, refine enrichment facts, adjust rule guards/parameters, and evolve the target profile.

\section{Case Study}
\label{sec:case_study}
We evaluate our approach on a long-lived \gls*{TOS} operated by a major European port infrastructure company. The system supports day-to-day terminal operations, including gate processing, yard management, cargo tracking, billing, and invoicing, and has been in continuous development since 2006. It is representative of the class of large, business-critical legacy applications we target: multi-site deployment, heterogeneous technology stacks, and extensive bespoke user interface and integration code. In this study, we scope the migration to the web tier and associated resources (e.g., interaction with the database), using the rest of the system as operational context.

The system is built on legacy versions of the .NET Framework and primarily consists of C\# code. The main web application uses ASP.NET MVC for the \gls*{UI} and coexists with a long-standing Web Forms layer, supported by an extensive library of bespoke components. Please note that, given the operational nature of the system, we report only the key technical aspects for security reasons.

\subsection{Core Layers and Architectural Components}
\begin{figure}[htbp]
    \centering
    \includegraphics[width=1\linewidth]{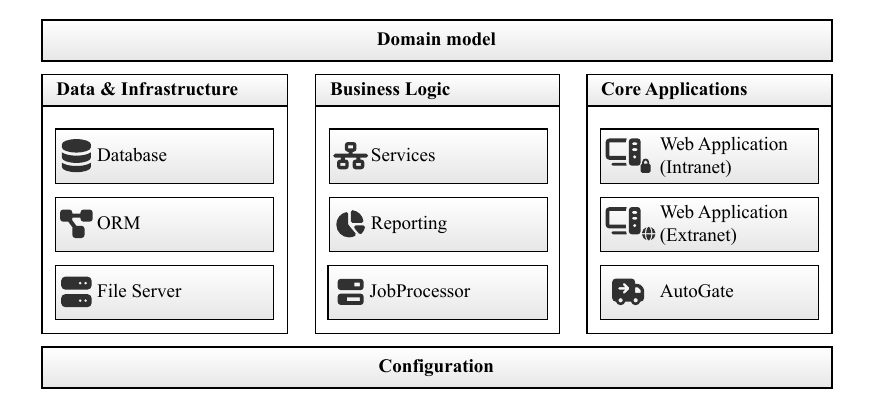}
    \Description[Diagram of Core layers and architectural components]{
    Layered architecture with a shared core (Configuration, Domain model); Data \& Infrastructure (Database, File Server, ORM); Business Logic (Services, Reporting, JobProcessor Service); and Applications (Core Applications: Web Application (Intranet), Web Application (Extranet), AutoGate).
    }
    \caption{Core layers and architectural components}
    \label{fig:core_system_architecture}
\end{figure}

The system follows a layered architecture (Fig.~\ref{fig:core_system_architecture}). At its center, the \emph{Domain Model} and \emph{Configuration} form a shared core that defines business concepts and runtime parameters across subsystems. 
The \emph{Data \& Infrastructure} layer provides platform services, comprising an Oracle database, a file server, and NHibernate-based \gls{ORM} for persistence. Above this, the \emph{Business Logic} layer implements cross-cutting functionality, including reporting, orchestration of application services, and background job execution via the JobProcessor component. The top \emph{Core Applications} layer exposes user-facing entry points: intranet and extranet web applications, the AutoGate module (driver-operated gate processing), and several smaller utilities (e.g., a test application). 

While this composition originally enforced a clear separation between infrastructure, business logic, and user-facing applications, years of evolution have introduced \emph{strong interdependencies between components}. Such coupling (e.g., shared libraries referenced across tiers and direct dependencies on infrastructure services) increases complexity and complicates modernization or stepwise migration. Additionally, although many architectural components have remained reliable and operational, they are increasingly \emph{constrained by diminishing tooling support and reduced compatibility} with modern environments.

\subsection{Challenges}
The system exhibits several \emph{intertwined challenges} that motivate modernization. Architecturally, tight coupling to business-critical dependencies constrains deployment, modular replacement, and refactoring, introducing operational risk. In parallel, core components depend on discontinued or outdated technologies with diminishing support.

Despite multi-site deployment, the reliance on hard-to-deploy, unsupported dependencies complicates rollout to new terminals and prolongs maintenance cycles. Knowledge erosion (i.e., fewer staff with deep system expertise) further complicates onboarding and raises the cost of change. 

Scale amplifies these issues: the web tier alone contains more than 1,500 web pages and 3\,GB of code, libraries, and documentation. Many artifacts are structurally similar yet diverge in details, driving duplication and inconsistency. Decades of incremental growth have produced architectural inefficiencies, reinforcing technical debt and impeding stepwise migration.

Collectively, operational stability has come at the expense of adaptability, maintainability, and responsiveness to changing business needs, making modernization necessary.

\section{Implementation and Results}
\label{sec:implementation_results}

\subsection{Analysis of the Current System}
As an initial step, we performed a structural analysis of the current-generation system to inventory source artifacts, codebase structure, architectural layering, and inter-component dependencies. The web tier is hybrid (ASP.NET Web Forms and ASP.NET MVC); our analysis and subsequent migration concentrate on the Web Forms-heavy UI. We combined targeted manual inspection with automated structural parsing to quantify relevant artifacts (UI components — both built-in ASP.NET and bespoke — declared properties, localization resources, and associated server-side logic).

Concretely, Web Forms pages consist of \texttt{.aspx} markup with \texttt{.aspx.cs} code-behind and \texttt{.designer.cs} files; reusable UI fragments are implemented as \texttt{.ascx} user controls; localized strings reside in \texttt{.resx} files. Our inventory identified about 1{,}500 \texttt{.aspx} pages, 1{,}100 \texttt{.aspx.cs} code-behind files, and 500 \texttt{.ascx} user controls. Across these pages, we observed around 30 distinct built-in ASP.NET \glspl*{UI} components (e.g., buttons, labels, text boxes), plus 700 bespoke \gls*{UI} components – each with tens of thousands of uses. The application further contains over 6{,}000 unique \texttt{.resx} strings bound to \gls*{UI} controls.

\begin{figure}[htbp]
    \centering
    \includegraphics[width=1\linewidth]{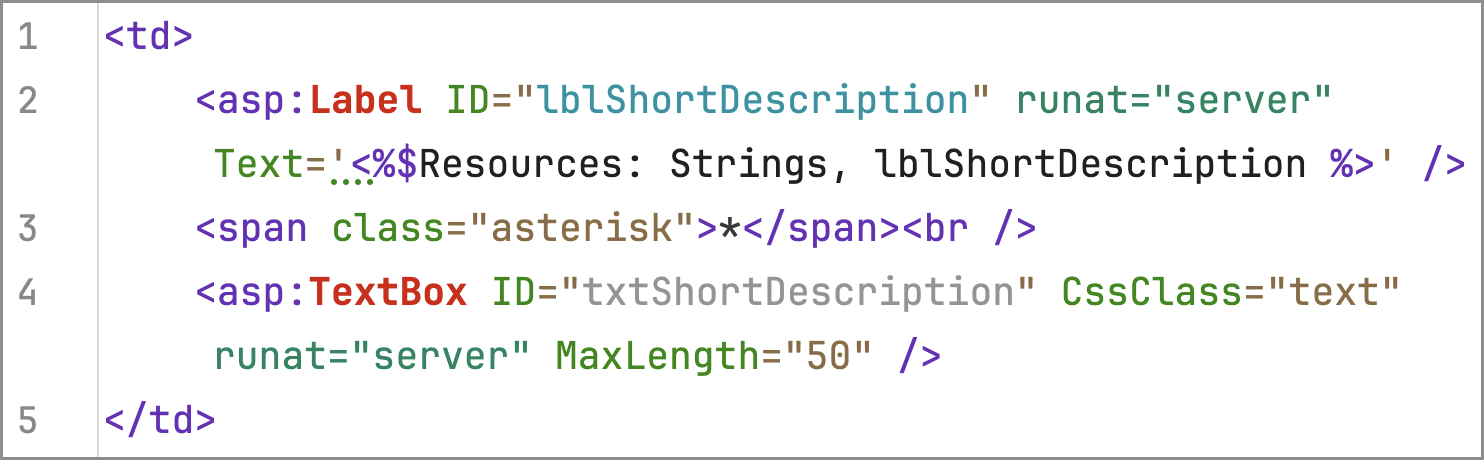}
    \Description[ASP.NET Web Forms Markup Example Code Excerpt]{
    Visible is an ASP.NET Web Forms code fragment. It defines a Label control with ID "lblShortDescription" whose Text property is bound to a localized resource key. Below that, a TextBox control with ID "txtShortDescription" is defined with a MaxLength property set to 50.
    }
    \caption{ASP.NET Web Forms markup, code excerpt (\texttt{.aspx})}
    \label{fig:code_snippet1}
\end{figure}

\subsection{Abstraction and Model Enrichment}
We implemented the intermediate model in Python to facilitate rapid parsing, rule authoring, and traceability. The model captures \emph{key architectural components} (e.g., Page, UserControl, Component, ResourceString) and associates them with \emph{semantic metadata} such as element location, typed relations (e.g., containment, depends-on, binds), and provenance links to the original artifacts. For each \emph{Page}, the model records its registered \emph{UserControls}, occurrences of built-in and bespoke \emph{Components}, and bound \emph{ResourceStrings}.

We enrich the model using \emph{Tree-sitter}~\cite{tree-sitter_2025} to obtain abstract syntax trees for C\# and markup, enabling consistent extraction of structural and semantic information (e.g., control hierarchies, property bindings, event handlers). This enrichment yields precise source-level detail sufficient to drive transformation and validation.

\subsection{System Synthesis and Transition}
We selected \emph{Next.js} as the target web framework for its mature ecosystem and modern capabilities (server-side rendering, static site generation, routing). Core cross-cutting features (authentication, navigation) were implemented manually and parameterized by the intermediate model.

Using the enriched model as an intermediate abstraction, we generated Next.js artifacts for all pages and for all uses of standard ASP.NET UI components via entirely manually defined transformation rules. Code generation was implemented with the Jinja2 templating engine, while page contents, hierarchy, naming, and ordering were automatically derived from the model. The transformation preserved the original page hierarchy, namespace structure, and file organization to facilitate traceability and comparison between legacy and target artifacts.

\begin{figure}[htbp]
    \centering
    \includegraphics[width=1\linewidth]{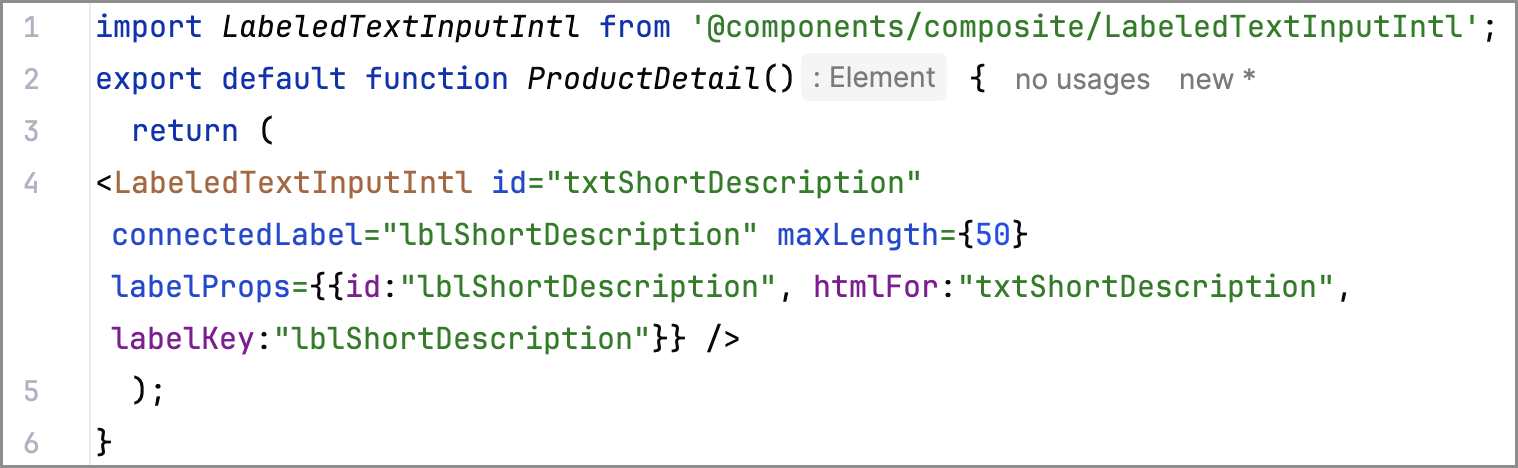}
    \Description[Next.js component generated from ASP.NET snippet]{React/Next.js component that corresponds to the ASP.NET example: a labeled text input with max length 50 and an internationalized label bound via a resource key.}
    \caption{Migrated Next.js page content, code excerpt (\texttt{.tsx})}
    \label{fig:code_snippet2}
\end{figure}

Fig.~\ref{fig:code_snippet1} shows an excerpt from a legacy \texttt{.aspx} page (label and text box, the label bound to a localized resource key). Fig.~\ref{fig:code_snippet2} presents the generated \texttt{.tsx} excerpt in Next.js, where the elements are realized as React components with preserved internationalization bindings. Equivalent transformations were applied to all covered ASP.NET components across the page set, maintaining functional behavior while enabling modernization of the web stack (manually verified).

\section{Evaluation}
\label{sec:evaluation}
We qualitatively assess the feasibility and outcomes of applying the proposed model-driven migration to the case system, focusing on maintainability, extensibility, and architectural quality. Evidence is drawn from generated artifacts, code reviews after migration, and developer feedback during adoption. Quantitative measurements are outside the scope of this experience report.

\paragraph{RQ1: Automation Potential}
The enriched intermediate model enabled automated extraction of pages, user controls, resource bindings, and standard UI components, followed by rule-based generation of target artifacts. Manual adaptation was only occasionally required and was largely confined to non-standard composites and custom layouts. \emph{Evidence:} complete generation for all pages and standard components; presence of diagnostic stubs for partial matches. \emph{Boundary:} patterns involving dynamic composition or bespoke layout logic resisted full automation.

\paragraph{RQ2: Implementation and Challenges}
Separating platform-in\-de\-pen\-dent abstractions from platform-specific logic proved essential to making rules reusable. Main challenges were  inconsistencies in legacy components, undocumented dependencies, and heavy use of custom UI controls alongside built-in ones. \emph{Evidence:} recurring partial matches flagged by guards, manual review of uncertain relations, and follow-up fixes driven by extractor heuristics. \emph{Outcome:} despite these issues, the intermediate model supported a systematic, repeatable transformation process, suggesting applicability to similar systems.

\paragraph{RQ3: Maintainability, Extensibility, and Architectural Quality}
We observed reduced duplication, clearer module boundaries, and easier onboarding due to familiar target idioms and preserved trace links.
\emph{Evidence:} code-review notes on module structure and reuse, developer reports of simpler navigation between legacy and generated artifacts, and fewer ad hoc conventions. \emph{Outcome:} While quantitative metrics were not collected, the qualitative results indicate an improvement in maintainability and extensibility.

\section{Limitations and Future Work}
\label{sec:limitations_future_work}
\paragraph{Limitations.}
Most built-in \glspl*{UI} components were migrated automatically; 
however, many  bespoke components used for layout composition or near-duplicate variants (same behavior, differing structure/styling) were not preserved verbatim. Patterns involving dynamic control creation, reflection-based wiring, or string-built queries also resisted static extraction and full automation. Our pipeline currently targets the web tier; data-access and domain-service transformations were only partially considered. The implementation is tuned to one target stack, and our extractors rely on markup grammars with incomplete coverage of edge constructs.

\paragraph{Threats to validity.}
\emph{Internal validity:} authors participated in design and evaluation, which can bias assessments; we mitigated this through code reviews and diagnostics, but quantitative checks were limited. \emph{Construct validity:} we emphasize qualitative indicators (maintainability, extensibility, developer experience) without standardized metrics, risking construct undercoverage. \emph{External validity:} results derive from a single system and a specific target stack; generalization to other domains, stacks, or organizational settings may vary. \emph{Conclusion validity:} absence of controlled baselines and statistical analysis limits the strength of comparative claims.

\paragraph{Future work.}
(i) Extend coverage to data-access and service layers, including transactional boundaries and consistency policies; (ii) generalize the target profile to additional stacks and deployment models; (iii) quantify outcomes with standardized measures (e.g., code churn, duplication, module coupling, onboarding time, change lead time) and non-functional metrics (latency, resource use); (iv) incorporate energy measurements across pre-/in-/post-migration phases; (v) explore LLM-assisted reconstruction of bespoke layouts and component variants with guardrails (deterministic scaffolds, diffable outputs, and invariant checks); and (vi) release a replication package (schemas, rules, and anonymized model snapshots) to enable independent evaluation and pattern reuse.

\section{Conclusions}
\label{sec:conclusions}
We presented a model-driven approach to modernizing large-scale legacy systems via an enriched, technology-agnostic intermediate model. Applied to an industrial system and manually verified, the approach automated substantial portions of the migration pipeline and preserved functional behavior as well as key non-functional qualities for standard components. The resulting target codebase benefits from improved maintainability and extensibility, supported by traceability between legacy artifacts and generated outputs. Challenges remain for bespoke layouts and tightly coupled logic, which require targeted manual adaptation. Overall, the evidence suggests that intermediate, model-based abstractions reduce risk and effort while enabling scalable, repeatable modernization. 
The work brings together a four-stage, end-to-end process, an intermediate model that makes structure, dependencies, and semantics explicit, and declarative, dependency-aware transformation rules that safeguard behavior and quality attributes. Applied and evaluated in an industrial setting, this combination automated large parts of the pipeline while clearly delineating the remaining manual adaptations, demonstrating both feasibility and transferability. 
Future work will extend coverage beyond the web tier and add quantitative evaluations.

\begin{acks}
This work was supported by ITEA4 and the Eureka Cluster on software innovation project "23016 GreenCode" supported with funding from the Federal Ministry of Research, Technology and Space (BMFTR), Germany (grant number 16IS24070G).
\end{acks}

\printbibliography

\appendix

\end{document}